\documentclass[aps,showpacs,prl,amssymb,amsmath,twocolumn, noshowpacs]{revtex4}
\usepackage{amsmath}
\usepackage{amsthm}
\usepackage{eucal}
\usepackage{amssymb}
\usepackage{mathrsfs}
\usepackage{graphicx}
\usepackage{amsmath,latexsym,amssymb}
\usepackage{graphicx}
\usepackage{mathtools}
\usepackage{latexsym}
\usepackage{amssymb}
\usepackage{bm}
\usepackage{color}
\usepackage{multirow}

\newcommand{\bra}[1]{\ensuremath{\langle #1 |}}
\newcommand{\ket}[1]{\ensuremath{| #1 \rangle}}

\begin{document}

\title{Detailed Balance of Thermalization dynamics in Rydberg atom quantum simulators}
\author{Hyosub Kim, Yeje Park, Kyungtae Kim, H.-S. Sim, and Jaewook Ahn}
\affiliation{Department of Physics, KAIST, Daejeon 34141, Korea}
\date{\today}

\begin{abstract}
Dynamics of large complex systems, such as relaxation towards equilibrium in classical statistical mechanics, often obeys a master equation~\cite{Pathria}. The equation significantly simplifies the complexities but describes essential information of occupation probabilities. 
A related fundamental question is the thermalization, a coherent evolution of an isolated many-body quantum state into a state that seems to be in thermal equilibrium. 
It is valuable to find an effective equation describing this complex dynamics.
Here, we experimentally investigate the question by observing sudden quench dynamics of quantum Ising-like models implemented in our 
quantum simulator, defect-free single-atom tweezers in conjunction with Rydberg atom interaction.
We find that saturation of local observables, a thermalization signature, obeys a master equation experimentally constructed by time-resolved monitoring the occupation probabilities of prequench states and imposing the principle of the detailed balance. 
Our experiment agrees with theories, and demonstrates 
the detailed balance in a thermalization dynamics that does not require coupling to baths or postulated randomness.
\end{abstract}
\maketitle
  
It is a long-standing question whether and how a closed many-body quantum system coherently evolves 
into a steady state~\cite{Polkovnikov2011,Eisert2015,Christian2016}.
As variety of quantum simulators have been developed recently, there are number of experimental reports on thermalization~\cite{Langen2013, Clos2016, Georg2016, Kaufman2016, Neill2016}.
There are also theoretical mechanisms, such as the eigenstate thermalization hypothesis (ETH)~\cite{Deutsch1991, Srednicki1994, Srednicki1999, Rigol2008, Rigol2012, D'Alessio2016},
 that tell us whether the steady state is practically indistinguishable from an equilibrium thermodynamic ensemble.

By contrast, the principles of dynamics toward the steady state remain largely unknown. The thermalization dynamics has the complexity exponentially increasing with system size; hence,
its computation is impractical for large systems.
Recently, a master equation was derived~\cite{Ates2012} for describing the thermalization dynamics of a quantum spin system.
It is constructed in terms of transition rates between the eigenstates of the prequench Hamiltonian, and well describes the time evolution of local observables towards steady-state values (excepts some coherent oscillations). 
It is powerful as the number of the rates necessary for the construction increases only linearly with system size.
We experimentally construct the master equation by preparing optical dipole traps with unit occupation of $^{87}$Rb atoms, and monitoring the global sudden quench dynamics to a Rydberg level.
 
\begin{figure}[tbh]
\centering
\includegraphics[width=0.45\textwidth]{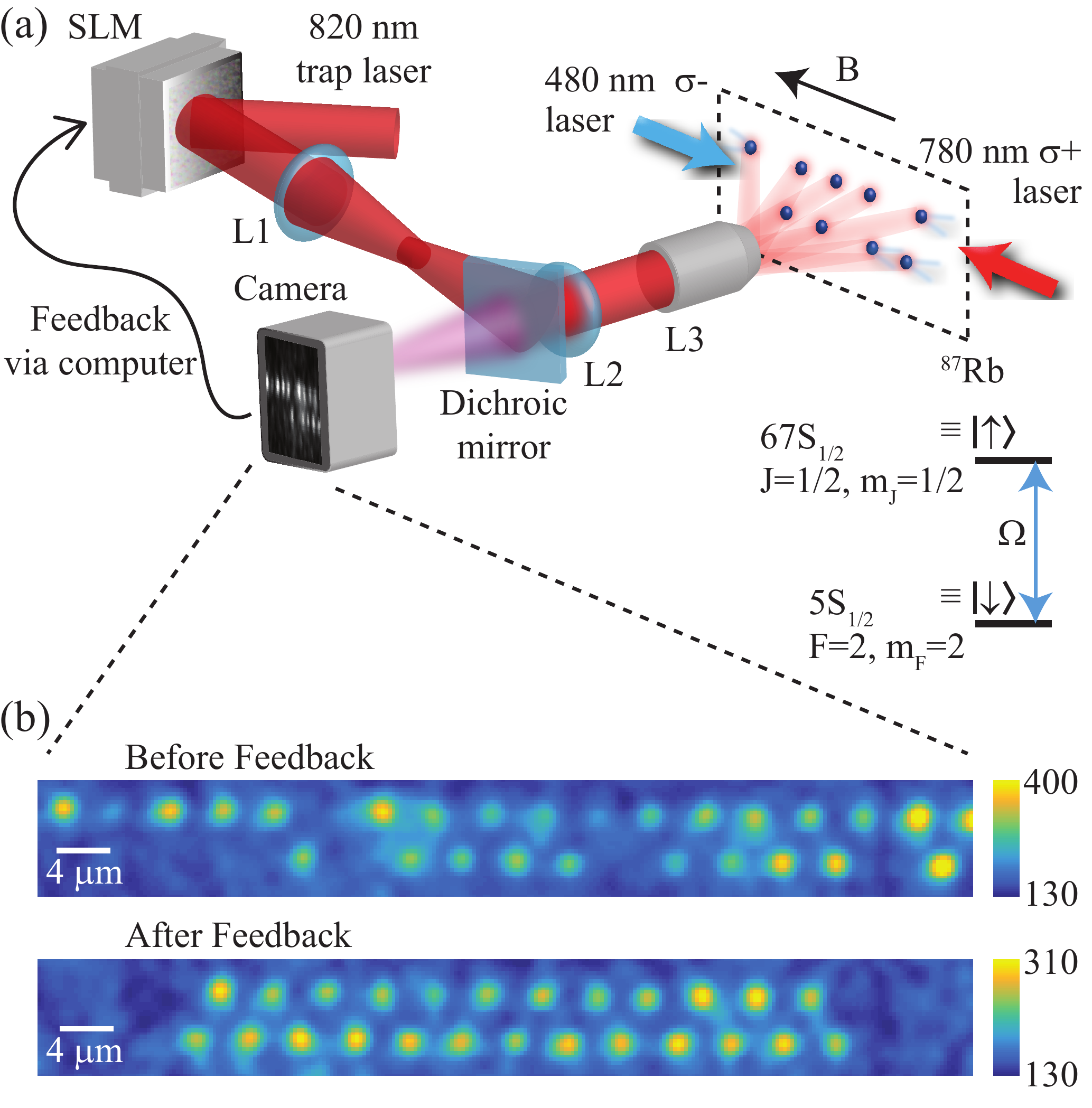} 
\caption{{\bf{Setup.}} (a) The wave-front of trap laser is modulated by a liquid-crystal spatial light modulator (SLM) and imaged by a telescope (L1, L2) and an objective lens (L3). 
The fluorescence image of the trapped atoms is captured by a camera, and analyzed to feedback the SLM for array compactification by atom shuttling. Then, 480~nm and 780~nm lasers excite the array to $67S$ Rydberg state, forming a tunable Ising-like spin-1/2 chain. 
(b) Zig-zag chain fluorescence images (Gaussian filtered for clarity).  }
\label{fig1}
\end{figure}

\noindent 
{\bf Rydberg atom experiment.}
\noindent  
We utilized the recently developed single-atom array synthesizer~\cite{Kim2016,Lee2017,Barredo2016,Endres2016} in conjunction with global Rydberg atom excitation. 
In Fig.~\ref{fig1}(a), defect-free $^{87}$Rb single-atom chains of various size $N=10-25$ were formed by using dynamic holographic optical tweezers; note the images of an $N=25$ zigzag chain of bending angle $\theta=60^\circ$ in Fig.~\ref{fig1}(b).
We fixed the interatom distance $d=4.0(2)~\mu$m and changed the zigzag angle $\theta$ from $45^\circ$ to $180^\circ$.
The entire array was coherently driven to $67S_{1/2}$ Rydberg state with homogeneous interaction strength by adopting widely used two-photon excitation~\cite{Low2012, Labuhn2016, Schaub2015, Bernien2017, Lienhard2017} (see Fig.~\ref{figs_levelscheme} in Supplements).
Each atom~$i$ behaved as a pseudo spin-1/2 system composed of the ground state $\ket{5S_{1/2}, F=2, m_F=2}\equiv\ket{\downarrow_i}$ and the Rydberg state $\ket{67S_{1/2}, J=1/2, m_J=1/2}\equiv\ket{\uparrow_i}$, as  intrinsic dephasing time $16~\mu$s was longer than  experiment duration $3~\mu$s (see Supplements).
   
The system can be described by the Hamiltonian of an Ising-like spin-1/2 chain~\cite{Labuhn2016, Schaub2015, Bernien2017, Lienhard2017}, 
\begin{equation}
H= H_0 + H_I=\sum_{i>j}V_{ij}\hat{n}_i\hat{n}_j + \sum_i \frac{\hbar\Omega}{2}\hat{\sigma}_x^{(i)}-\frac{\hbar \Delta}{2}\hat{\sigma}_z^{(i)},
\label{IsingModel}
\end{equation}
where $\hat{n}_i=\ket{\uparrow_i}\bra{\uparrow_i}$, $\hat{\sigma}_x^{(i)}=\ket{\uparrow_i}\bra{\downarrow_i}+\ket{\downarrow_i}\bra{\uparrow_i}$, and $\hat{\sigma}_z^{(i)}=\ket{\uparrow_i}\bra{\uparrow_i}-\ket{\downarrow_i}\bra{\downarrow_i}$.  
The first term $H_0$ of Eq.~\eqref{IsingModel}, the repulsive van der Waals interaction $V_{ij}=-C_6/|\bm{x}_i-\bm{x}_j|^6$
with coefficient $C_6=-470$~GHz/$\mu$m$^6$~\cite{Sebastian2017}, behaves as interactions between the spins, while the second and third terms with Rabi frequency $\Omega$ and detuning $\Delta$ act as spin transverse Zeeman splitting $H_I$.
The nearest neighbor interaction strength is estimated as $V_{12}/2\pi \hbar=14-25$~MHz for $d=4.0(2)~\mu$m, 
and the next nearest neighbor interaction strength $V_{13}$ depends on $\theta$;
$V_{13} = V_{12}/64 $ for $\theta = 180^\circ$ and
$V_{13} = V_{12} $ for $\theta = 60^\circ$.
The fluctuation of $V_{ij}$ is due to thermal atomic motions.
There is a shot-to-shot noise on $H_I$, resulting in $2~\mu$s inhomogeneous dephasing time on collective Rabi oscillations (see Fig.~\ref{figs_RyfractionData} in Supplements); however, it does not qualitatively alter the equilibration dynamics under the parameters of  $H_0 \gg \hbar \Omega \gg \hbar |\Delta|$~\cite{Lesanovsky2010}.

\begin{figure}[tbh]
\centering
\includegraphics[width=0.48\textwidth]{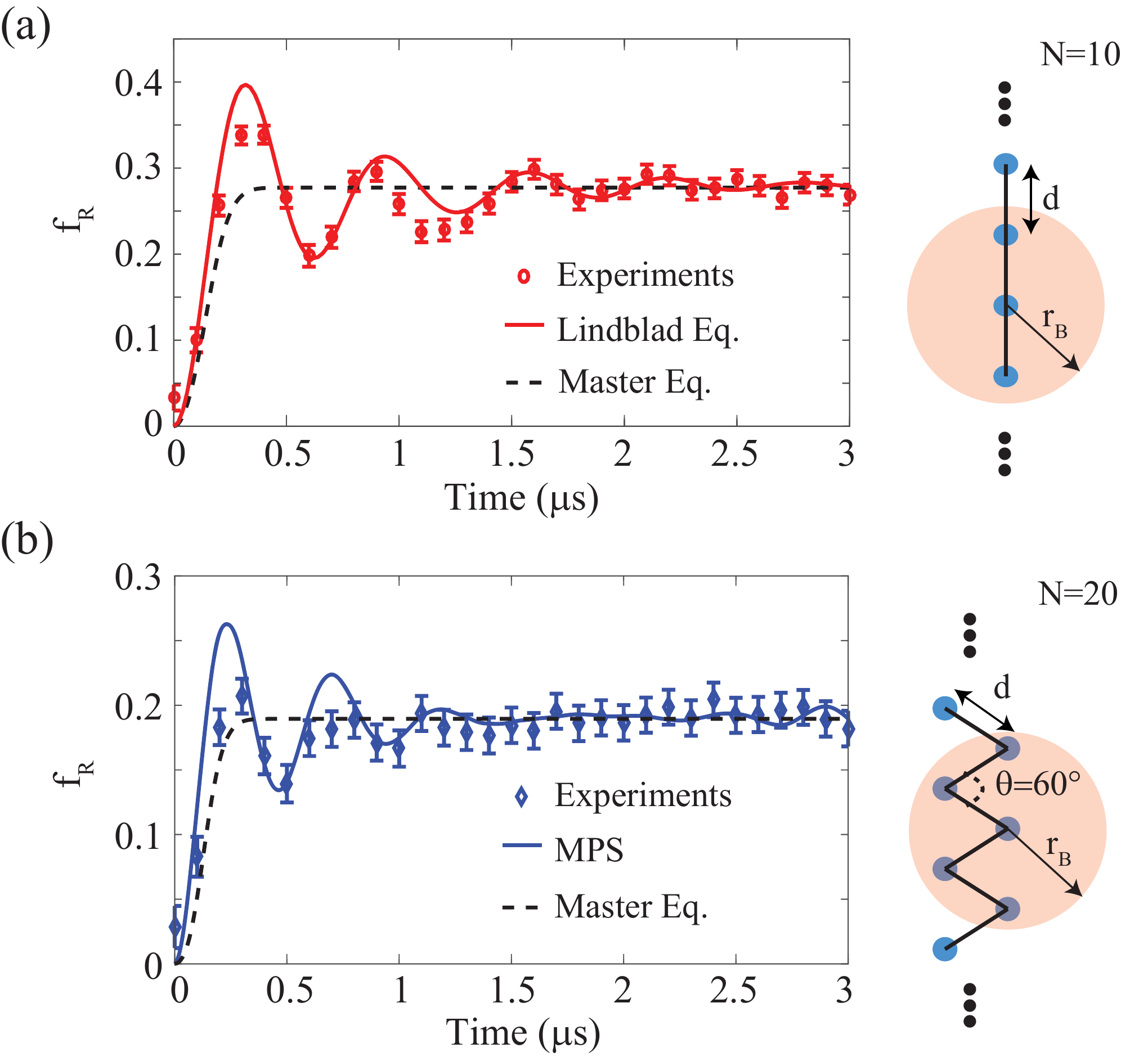} 
\caption{{\bf{Thermalization dynamics.}} Time dependence of Rydberg fraction after the quench for (a) the linear chain of $N=10$ and $\theta = 180^\circ$ and (b) the zigzag chain of $N=20$ and $\theta = 60^\circ$.
The experimental data (circles) are compared with the computation (solid lines)  and the result (dashed) of the master equation constructed based on the experimental data.
The errorbars are standard error of the mean.
Right panels: Chain configurations with blockade radius of $r_B={(|C_6|/2\pi\Omega)}^{1/6}=6.5~\mu$m and lattice spacing of $d=4.0(2)~\mu$m.}
\label{fig2}
\end{figure}

\vspace{.3cm}
\noindent 
{\bf Thermalization.} 
Initially, $\Omega$ and $\Delta$ are zero and the chain is in the ground state $|\downarrow_1 \downarrow_2 \cdots \downarrow_N \rangle$ of $H_0$. Then, saying at $t=0$, they are suddenly turned onto  $\Omega/2\pi=1.0(1)$~MHz and $\Delta/2\pi=0.0(1)$~MHz. 
This global quench makes the measured Rydberg fraction $f_R \equiv\sum_i\langle \hat{n}_i \rangle (t)/N$ change in time as in Fig.~\ref{fig2} (also see $\sum_{i,j} \langle \hat{n}_i  \hat{n}_j \rangle (t)/N^2$ in Fig.~\ref{figsNsquaredata} of Supplements).
The overall features of $f_R$ are qualitatively the same for $N \geq 10$; $f_R$ shows coherent oscillations before it approaches to a steady-state value $\bar{f}_R$.
The major frequency component of the oscillations occurs at $\sqrt{2} \Omega$ for the linear chain and $\sqrt{3} \Omega$ for the zigzag chain of $\theta = 60^\circ$, corresponding to the collective Rabi frequency of two or three atoms. 
The results agree with computations based on a Lindblad equation for $N=10$
and on matrix product states (MPS) for $N=20$ (see Supplements).
Note that the shot-to-shot noise is taken into account for $N=10$; the noise effect becomes negligible for larger $N$~\cite{Clos2016, Rigol2008, Kiendl2017}.

Around the relaxation time $t_\textrm{relax} = 1.5 - 2$~$\mu$s for $\theta=60^\circ - 180^\circ$, the oscillations are suppressed. We obtain  the time-average $\bar{f}_R$ at $t\geq t_\textrm{relax}$.
$\bar{f}_R$ follows the universal scaling behavior of $\bar{f}_R \propto \alpha^\nu$ with $\alpha \propto \hbar \Omega d^6 / |C_6|$ (see Fig.~\ref{figsCriticalBehaviors} in Supplements). The measured exponent $\nu = 0.16(2)$ agrees with the prediction~\cite{Low2012, Weimer2008} based on the Hamiltonian $H$.
All the above observations support that our system properly works as Rydberg quantum simulators~\cite{Labuhn2016, Bernien2017, Lienhard2017, Marcuzzi2017,Guardado2017}. 
 
 \begin{figure*}[tbh]
\centering
\includegraphics[width=0.96\textwidth]{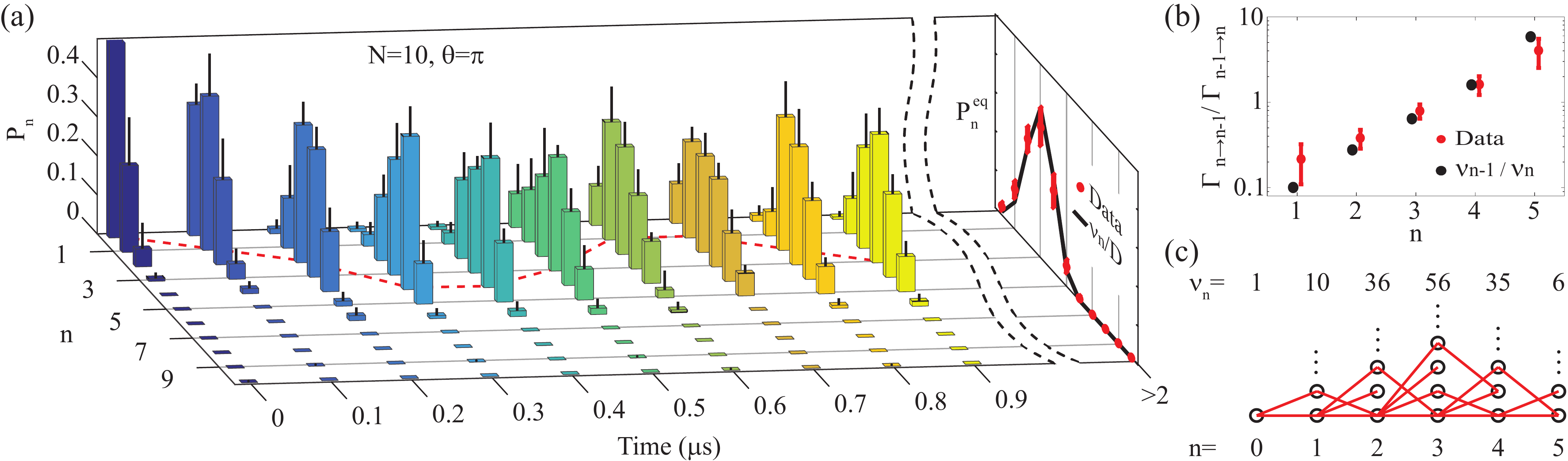} 
\caption{{\bf{Detailed balance in the linear chain of $\bm{N=10}$.}} (a) The measured $P_n(t)$ (color bars) and its standard deviation ($\pm\sigma$, overlaid black lines).
At $t > 2$~$\mu$s, the measured data (red dots) and the theoretical predictions (black line)  of the steady-state values $P^\textrm{eq}_n$ are shown.
The red dashed line evolves along the tallest bar, showing the coherent oscillation of $P_n$.
(b) The ratio $\Gamma_{n\rightarrow n-1}/\Gamma_{n-1\rightarrow n}$ (red dots) of transition rates are obtained (with errorbar $\pm\sigma$) from $P^\textrm{eq}_n$, and compared with the theoretical prediction (black dots).
(c) Graph for the thermalization dynamics. 
Its nodes (circles) represent 
low-energy eigenstates of the prequench Hamiltonain $H_0$, classified by the number $n$ of spin-up atoms in the states. $\nu_n$ is the number of the eigenstates having $n$ spin-up atoms.
Each red link connecting two nodes indicates transitions between the corresponding states by $H_I$. 
}
\label{fig3T}
\end{figure*}

 \vspace{.3cm}
\noindent 
{\bf Detailed balance.} 
To analyze the relaxation of $f_R$, we measure the probabilities $P_n(t)$ with which there are $n$ atoms in spin up $\ket{\uparrow}$ at time $t$. 
In Fig.~\ref{fig3T}(a), $P_n(t)$ exhibits coherent oscillations while diffuses to a steady-state distribution.
To describe the diffusion, we consider a master equation of the simplest form~\cite{Pathria,Ates2012} 
\begin{align}
\partial_t P_n(t) & = [ P_{n+1}(t) \Gamma_{n+1 \to n} (t) + P_{n-1}(t) \Gamma_{n-1 \to n} (t) ]\nonumber\\
& - P_n(t) [\Gamma_{n \to n-1} (t) + \Gamma_{n \to n+1} (t)], \label{master}
\end{align}
where $\Gamma_{n \to n \pm 1}(t)$ is the rate of transition from states with $n$ spin-up atoms to those with $n \pm 1$.
The other transitions of $n \leftrightarrow n' (\ne n \pm 1)$ are negligible in our regime of $H_0 \gg H_I$, as they are higher-order processes of multiple spin flips by $H_I$. 

The principle of the detailed balance, $[ P^\textrm{eq}_{n+1} \Gamma_{n+1 \to n} + P^\textrm{eq}_{n-1} \Gamma_{n-1 \to n} ]
= P^\textrm{eq}_n [\Gamma_{n \to n-1} + \Gamma_{n \to n+1}]$ and $P^\textrm{eq}_{1} \Gamma_{1 \to 0} 
= P^\textrm{eq}_0 \Gamma_{0 \to 1}$, is obtained from the master equation in the steady state where $P_n = P^\textrm{eq}_n$ and $\partial_t P_n(t)  = 0$; equivalently, $\Gamma_{n\rightarrow n-1}/\Gamma_{n-1\rightarrow n}=P^\textrm{eq}_{n-1}/P^\textrm{eq}_n$.
We obtain $P^\textrm{eq}_{n}$ by the time average of $P_n(t)$ at $t \geq t_\textrm{relax}$,
and retrieve the microscopic information of 
$\Gamma_{n\rightarrow n-1}/\Gamma_{n-1\rightarrow n}$, using the detailed balance relation.
In Fig.~\ref{fig3T}, the results agree with the theoretical prediction~\cite{Ates2012,Olmos2010} of
$P^\textrm{eq}_n=\nu_n/D$ and $\Gamma_{n\rightarrow n-1}/\Gamma_{n-1\rightarrow n}= \nu_{n-1}/\nu_n$ obtained in the limit of $H_I/H_0 \to 0$, where $\nu_n={N+1-n \choose n}$ for the linear chain,  $\nu_n = {N+2-2n \choose n}$ for $\theta = 60^\circ$, and $D = \sum_n \nu_n$~\cite{Olmos2010}.

We explain the meaning of $\nu_n$ for the linear chains as an example. 
In our relaxation dynamics, $H_0 \gg H_I$ and the initial state is the ground state $|\downarrow_1 \downarrow_2 \cdots \downarrow_N \rangle$ of $H_0$. In this case, it is enough to consider only low-energy eigenstates $|\sigma_z^{(1)} \sigma_z^{(2)} \cdots \sigma_z^{(N)} \rangle$ of the prequench Hamiltonian $H_0$, from $\ket{\downarrow_1 \downarrow_2 \downarrow_3 \cdots}$ to $\ket{\uparrow_1 \downarrow_2 \uparrow_3 \cdots}$,  in which any two neighboring spins $\sigma_z^{(i)}$ and $\sigma_z^{(i+1)}$ cannot be simultaneously in spin up;
the other higher-energy eigenstates can be ignored, since they are separated from the low-energy states in energy at least by $V_{i,i+1}$. Then, $P_n(t)$ almost equals the probability of occupying the low-energy states of $n$ spin-up atoms, and the possible values of $n$ are $0,1, \cdots, n_\textrm{max}=N/2$ for even $N$ and $0,1, \cdots, n_\textrm{max}=(N+1)/2$ for odd $N$.
 $\nu_n$ is the number of the low-energy states of $n$ spin-up atoms.
 Transitions between those of $n$ and those of $n\pm 1$, occuring with a single spin flip by $H_I$, govern the relaxation dynamics in our regime of $H_0\gg H_I$ as in Fig.~\ref{fig3T}(c).
 In this case, the ratio $\Gamma_{n\rightarrow n-1}/\Gamma_{n-1\rightarrow n}$ of the transition rates equals the ratio $\nu_{n-1}/\nu_n$.
We emphasize that the ratios, microscopic information of the dynamics, are measured in our experiments.

The master equation in Eq.~\eqref{master} efficiently describes the relaxation dynamics, as it has only $2n_\textrm{max}$ parameters of the transition rates $\Gamma_{n \to n \pm 1}$, which is much smaller than the size $2^N$ of the Hilbert space.
This allows us to experimentally construct the master equation.
Among the $2n_\textrm{max}$ parameters, $n_\textrm{max}$ parameters are determined by the ratios $\Gamma_{n\rightarrow n-1}/\Gamma_{n-1\rightarrow n}$ measured applying the detailed balance.
The other $n_\textrm{max}$ parameters are determined by using the probabilities $P_n(t)$ and their derivatives $\partial_t P_n(t)$ measured at the early stage of $t \simeq 0$ before the coherent oscillations occur (see Supplements). 
In this step, we use the form of $\Gamma_{n \to n \pm 1} (t) = 2 \Omega^2 t T_{n \to n \pm 1}$ derived in Ref.~\cite{Ates2012}, where $T_{n \to n \pm 1}$'s are time independent.
Using the experimentally contructed master equation, we compute the time evolution of Rydberg fraction $ f_R (t)$ ($ = \sum_n n P_n (t)/N$) and  $\sum_{i,j} \langle \hat{n}_i  \hat{n}_j \rangle (t) /N^2$ ($= \sum_n n^2 P_n(t)/N^2$)
and find that the result well describes the experimental data of the relaxation of $f_R(t)$ in Fig.~\ref{fig2} (see also Fig.~\ref{figsNsquaredata}(a) in Supplements).
Note that the master equation result does not show the coherent oscillations, since the higher-energy eigenstates and the processes of multiple spin flips are ignored in the master equation.
All the observations imply that the thermalization dynamics obeys the master equation, similarly to dynamics to equilibrium in statistical mechanics.

\begin{figure}[tbh]
\centering
\includegraphics[width=0.45\textwidth]{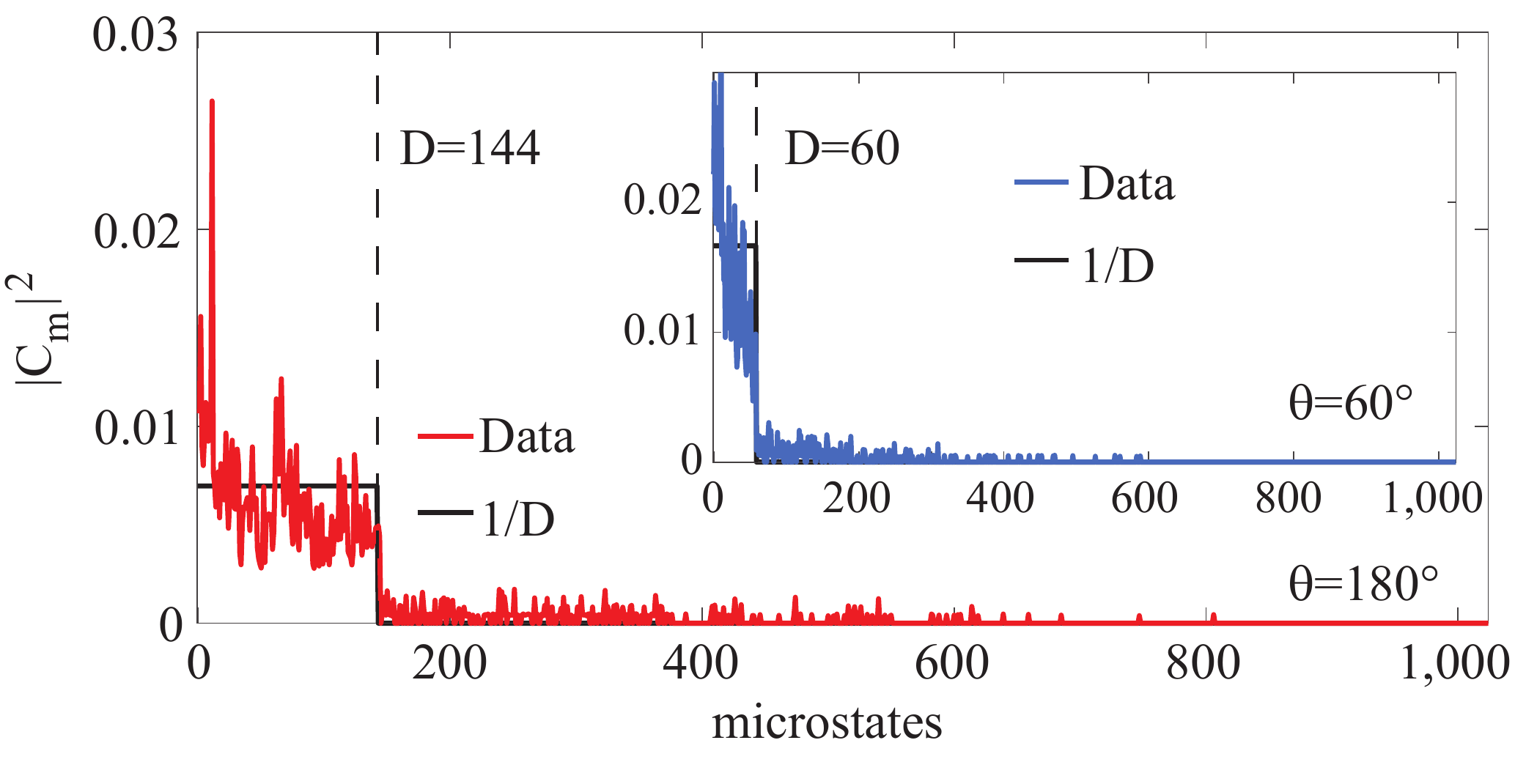} 
\caption{{\bf{Diffusion over prequench states. }} 
Time average ($t \geq t_\textrm{relax}$) of the measured probability $|C_m|^2$ of occupying the $m$-th eigenstates of the prequench Hamiltonian $H_0$ for the linear chain of $N=10$. The position of the $D$-th eigenstate is indicated by the dashed line. Inset: The same for the $\theta = 60^\circ$ zigzag chain of $N=10$.
}
\label{fig4MicrostatesAngle}
\end{figure}

\vspace{.3cm}
\noindent 
{\bf Steady State.} 
The thermalization dynamics can be considered as diffusion on the graph in Fig.~\ref{fig3T}(c), where each link has equal transition probability determined by $H_I$. This indicates that the relaxation time $t_\textrm{relax}$ depends on the initial point of the diffusion~\cite{Ates2012,Olmos2010}.
The initial state $|\downarrow_1 \downarrow_2 \cdots \downarrow_N \rangle$ of this experiment is located at  an edge of the graph. Hence the dynamics has a long relaxation time  $t_\textrm{relax}$  as in Figs.~\ref{fig2} and~\ref{fig3T}(a).
When an initial state is located closer to the center of the graph, the resulting coherent oscillations become more rapidly suppressed with shorter  $t_\textrm{relax}$~\cite{Ates2012}.

We experimentally measured the steady-state values of the probability $|C_m|^2$ with which the chain occupied the $m$-th eigenstate of the prequench Hamiltonian $H_0$. 
In the language of the diffusion on the graph, $|C_m|^2$ is interpreted as the occupation probability of the node corresponding to the $m$-th eigenstate. 
As shown in Fig.~\ref{fig4MicrostatesAngle}, the result is $|C_m|^2 \simeq 1/D$ , where $D = \sum_n \nu_n$ is the total number of the low-energy eigenstates.
This demonstrates almost uniform spreading over the graph, namely over the low-energy eigenstates.
 Indeed, the experimental data of $P^\textrm{eq}_n$ are close to $\nu_n / D$.
 
\begin{figure}[tbh]
\centering
\includegraphics[width=0.96\columnwidth]{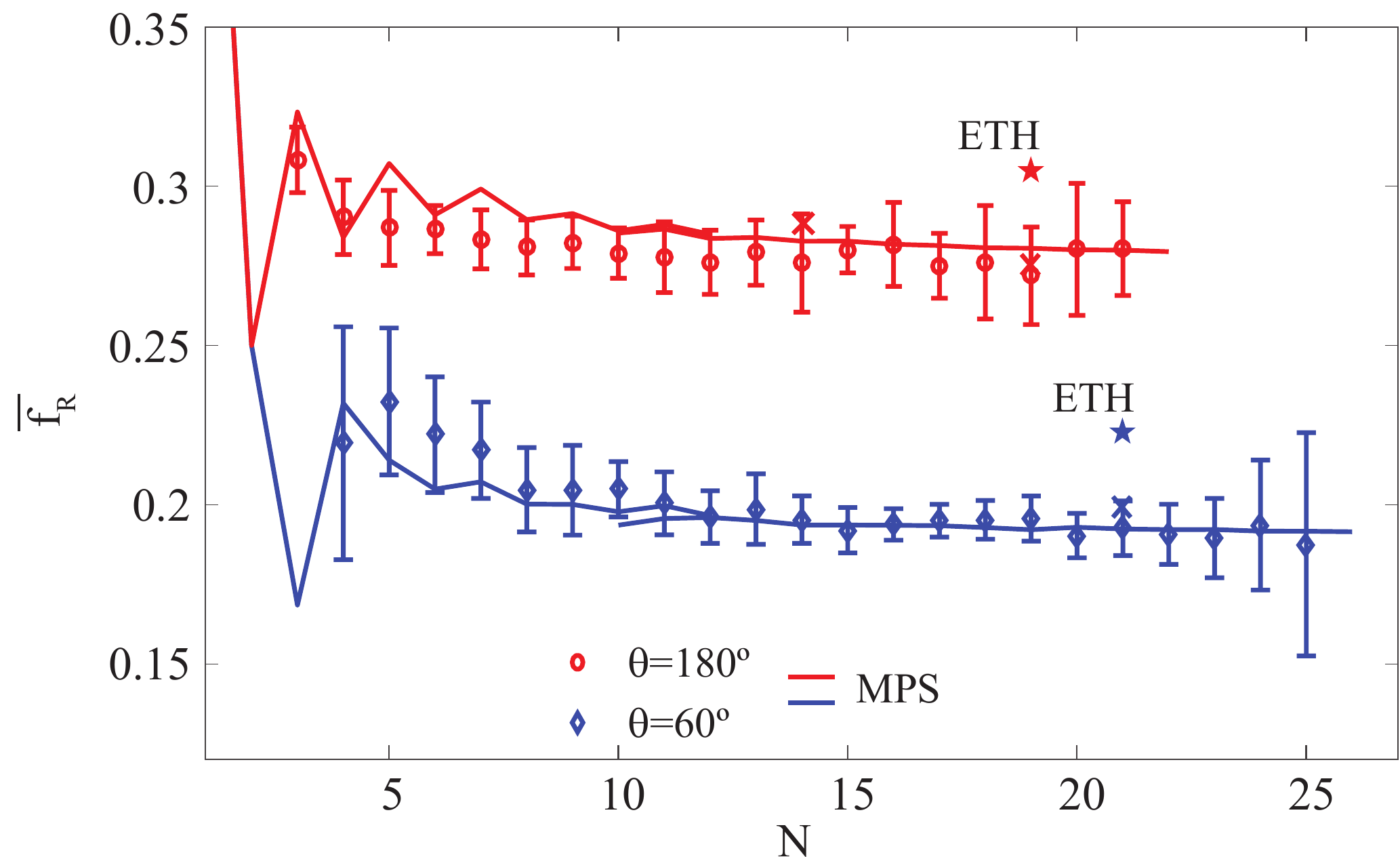} 
\caption{{\bf{Thermalization value.}} 
 Time average $\bar{f}_\textrm{R}$ of measured Rydberg fraction 
as a function of system size $N$ for the linear chain (red circles) and the zigzag chain of $\theta = 60^\circ$ (blue diamonds). 
The errorbars of $\pm \sigma$ are shown.
For comparison, { the MPS result (solid lines) 
and the ETH prediction (at $N=19$ for the linear chain and at $N=21$ for the zigzag chain; color stars) are shown.}  
}
\label{fig5NvsRyfraction}
\end{figure}

 In Fig.~\ref{fig5NvsRyfraction}, the experimental results of the steady-state values of the Rydberg fraction $f_\textrm{R}$ 
  are shown for $N = 3 - 25$.
They agree with the computation based on the MPS. 
They are however slightly different from the ETH prediction. Indeed, the typical features of the ETH do not hold in our cases (see Supplements).


\vspace{.3cm}
\noindent 
{\bf Conclusion.} 
In summary, we performed a quantum simulation experiment with tunable tweezer traps and Rydberg atom interaction. Our quantum simulator provides an ideal test bed for studying quantum coherent evolution of a many-body system after a quench. It allows to simulate a one-dimensional or two-dimensional lattice of Ising-like spin-1/2 particles or the Hamiltonian in Eq.~\eqref{IsingModel} with parameters tunable in a wide range. We can monitor the time evolution by measuring occupation probabilities of the eigenstates of a prequench Hamiltonian or a postquench Hamiltonian. The thermalization dynamics studied in our experiment belongs to the cases where the postquench Hamiltonian is slightly modified after quench so that $H_0 \gg H_I$.
Our results suggest that the detailed balance can be an underlying principle of the thermalization dynamics of the cases. The thermalization dynamics can be efficiently described by the diffusion governed by a master equation of a simple form, similarly to relaxation towards equilibrium in classical statistical mechanics but without its underlying assumptions of coupling to baths and the ergodicity hypothesis based on randomness.
   
\noindent 
\section{Supplements} 

\subsection{Experimental details} 
\begin{figure}[tbh]
\includegraphics[width=0.47\textwidth]{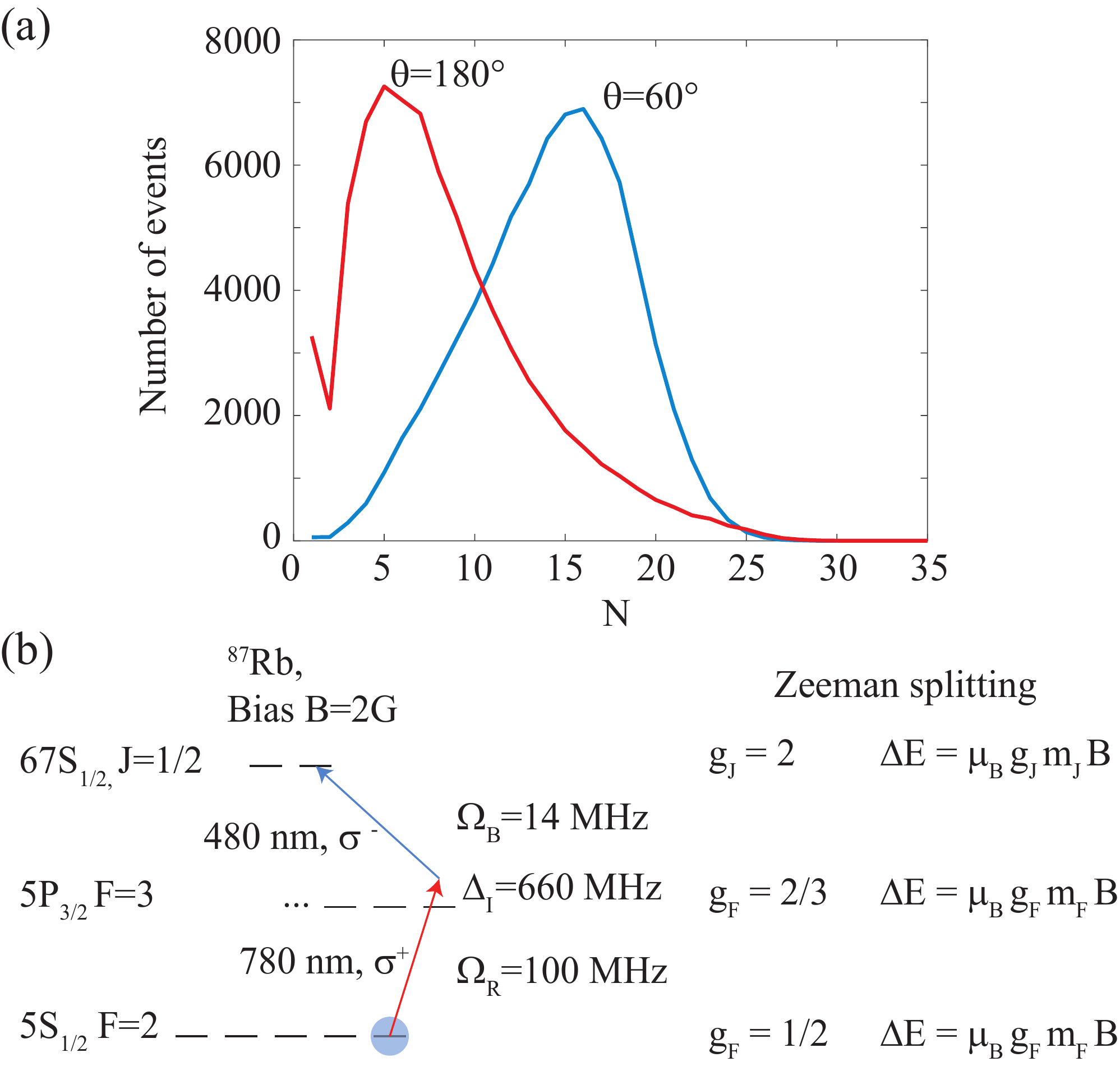} 
\caption{(a) Defect-free atom loading spectra, resulting from two-time of feed-back of total 35 sites, for $\theta=180^\circ$ (red) and $60^\circ$ (blue). (b) Graphical representation of our Rydberg excitation scheme with Rabi frequency and detuning (1/$2\pi$ scaled). The Rydberg hyper-fine splitting is sub-kHz order and thus treated in a reduced form. Zeeman splittings of individual states are shown as well, where the two-level scheme ($\ket{5S_{1/2}, F=2, m_F=2}$ and $\ket{67S_{1/2}, J=1/2, m_J=1/2}$) is robust against magnetic field fluctuation.}
\label{figs_levelscheme}
\end{figure}

The zigzag or linear arrays of $^{87}$Rb atoms were prepared by holographic tweezer traps~\cite{Kim2016,Lee2017}. The dipole-trap laser (820~nm) was phase-modulated with the SLM to form the zigzag chain of focused gaussian beams of waist $w_0=1$~$\mu$m, lattice constant $d=3.8-4.2$~$\mu$m, and bending angle $\theta$. Inside the chamber, a 3D magneto-optical trap (MOT) was overlapped with the dipole traps to form a partially filled ($p\approx 0.5$) single-atom array of 80~$\mu$K temperature. Then the camera took an image of the initial single-atom loading by collecting scattered photons from the same MOT beams for 40~ms. The result was fed back to the tweezer traps to produce a defect-free single-atom array through atom shuttling~\cite{Kim2016,Lee2017,Barredo2016,Endres2016}. The result was then measured and fed back once again for more successful array preparation. The final result showed defect-free array spectrum spanning wide range of atom numbers as shown in Fig.~\ref{figs_levelscheme}(a).

The initialized array was collectively and resonantly driven to Rydberg state $\ket{67_{1/2}, J=1/2, m_J=1/2}\equiv\ket{\uparrow}$ by using the release and recapture (R$\&$R) protocol and the counter-propagating two-pulses ($\sigma^+$ 780~nm and $\sigma^-$ 480~nm) scheme, as shown in Fig.~\ref{figs_levelscheme}(b) \cite{Low2012,Schaub2015,Labuhn2016,Bernien2017,Lienhard2017}.
First, the bias B-field of 2~G was turned on 
and waited 50~ms for eddy current decay from nearby metals, and the dipole-trap beam power was adiabatically reduced to 1/3 to mitigate AC stark shift. Then, the entire array was optically pumped to the ground state $\ket{5S_{1/2}, F=2, m_F=2}\equiv\ket{\downarrow}$ by using $\sigma^+$ transition of $\ket{5S_{1/2}, F=2}\rightarrow\ket{5P_{3/2}, F'=2}$ for 500~$\mu$s while the repump was kept low.  
After driving the system with the R$\&$R protocol, the dipole trap beam power was adiabatically recovered, the bias B-field was turned off, and waited for 50~ms again. Then the final single-atom image was captured for the site-resolved number state measurement, e.g. $\hat{n}\ket{\downarrow\uparrow}=01\ket{\downarrow\uparrow}$. The overall sequence repeated indefinitely at 0.6~Hz rate.

We used external-cavity diode lasers (780~nm and 480~nm) for Rydberg state excitation. The laser frequencies were locked to an ultra-low-expansion (ULE) cavity (Stable Laser Systems, ATF-6010-4),  
resulting in absolute frequency drift below 1~kHz on both lasers. The achieved linewidths were estimated to below 10~kHz. The intermediate level detuning from 5S$_{1/2}, F=2$ to 5P$_{3/2}, F=3$ was $\Delta_{I}=2\pi\times 660$~MHz. The 780~nm Rabi frequency, $\Omega_{780}=2\pi\times 100$~MHz, was calibrated by Stark shift measurement~\cite{Maller2015}. Then, the 480~nm Rabi frequency, $\Omega_{480}=2\pi\times 14$~MHz, was deduced by two-photon Rabi frequency, $\Omega_{780}\Omega_{480}/2\Delta_{I}=2\pi\times 1$~MHz.

\begin{figure}[tbh]
\includegraphics[width=0.48\textwidth]{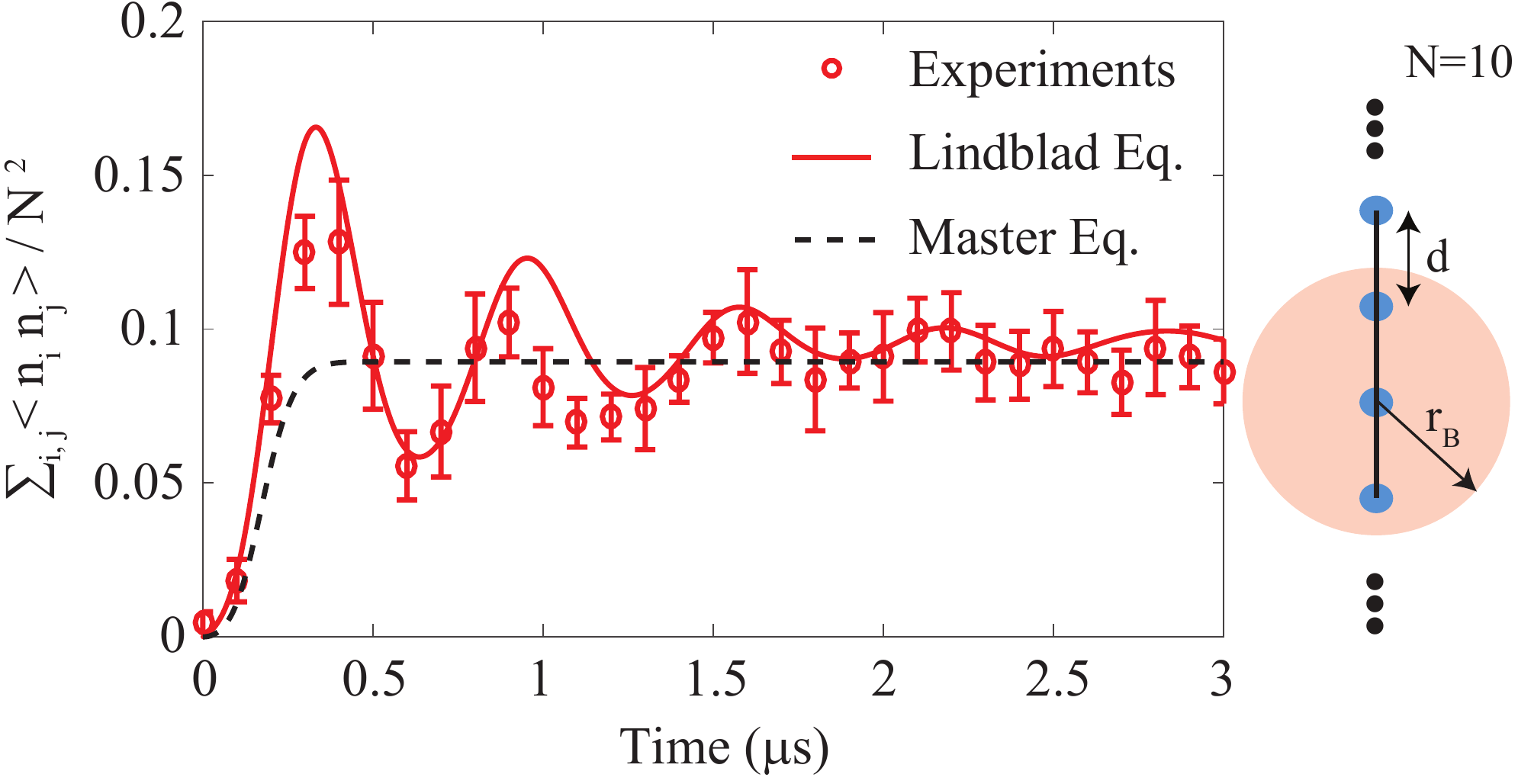} 
\caption{Time dependence of $\sum_{i,j} \langle \hat{n}_i  \hat{n}_j \rangle (t) /N^2$ measured for the linear chain of $N=10$.  
The experimental data (circles) are compared with the computation (solid lines) based on the Lindblad equation  and also with the result (dashed) of the master equation constructed based on the experimental data.
The errorbars are standard error of the mean.
} \label{figsNsquaredata}
\end{figure}

\subsection{Supporting experimental data: Thermalization}

\begin{figure}[tbh]
\includegraphics[width=0.46\textwidth]{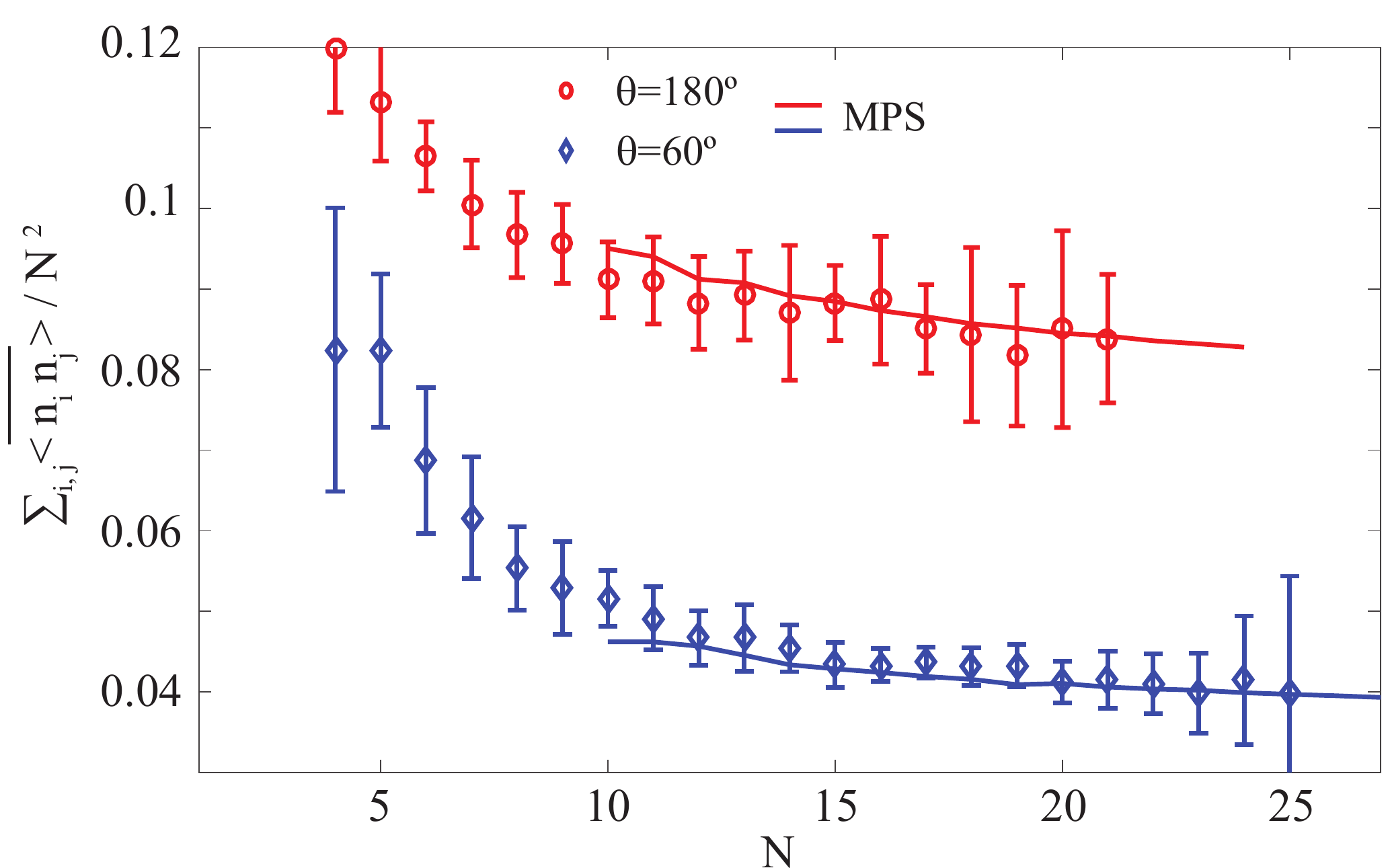} 
\caption{ 
 Time average  of  $\sum_{i,j} \langle \hat{n}_i  \hat{n}_j \rangle (t) /N^2$ at $t \ge t_\textrm{relax}$ 
as a function of system size $N$ for the linear chain (red circles) and the zigzag chain of $\theta = 60^\circ$ (blue diamonds). 
The errorbars of $\pm \sigma$ are shown. 
The theoretical results (solid lines) based on MPS are also shown.
}
\label{figNsquarevsN}
\end{figure}

In the main text, we have shown the experimental data of the Rydberg fraction $f_\textrm{R}$. Here we show the data of another observable of  $\sum_{i,j} \langle \hat{n}_i  \hat{n}_j \rangle (t)/N^2$. 
We plot the quench dynamics of $\sum_{i,j} \langle \hat{n}_i  \hat{n}_j \rangle (t)/N^2$ in Fig.~\ref{figsNsquaredata}.
Similarly to the dynamics of $f_\textrm{R}$ in Fig.~\ref{fig2}, the experimental data agree with the calculation based on the Lindblad equation and also with the master equation constructed from our experimental data.
This supports that our system is described by the Hamiltonian $H$.
We also plot the steady-state value of  $\sum_{i,j} \langle \hat{n}_i  \hat{n}_j \rangle (t)/N^2$ in Fig.~\ref{figNsquarevsN}.
The experimental results agree with the MPS calculation. 

\subsection{Supporting experimental data: Scaling behavior}
\begin{table*}[tbh]
\centering
\caption{Dephasing sources on individual atoms}
\label{T.individualdephasing}
\begin{tabular}{cccc}
\hline
\multicolumn{2}{c}{Error sources} & Effective Rabi decay $\tau$ & Treatment \\ \hline
\multicolumn{2}{c}{\multirow{3}{*}{\begin{tabular}[c]{@{}c@{}}5P$_{3/2}$ decay, 26 ns\\ 67S$_{1/2}$ decay, 100 $\mu$s, Ref.~\cite{MethodsBeterov}\\ $\Omega_1$, $\Omega_2$ imbalance\end{tabular}}} & \multirow{3}{*}{$\simeq$15.5 $\mu$s} & \multirow{3}{*}{\begin{tabular}[c]{@{}c@{}}  $L_i=\sqrt{\gamma/2}\sigma_z^i$, \\ where $\gamma=2\pi\times 20$ kHz\end{tabular}} \\
\multicolumn{2}{c}{} &  &  \\
\multicolumn{2}{c}{} &  &  \\ \hline
\multicolumn{2}{c}{5P$_{3/2}$ geometric phase} & \textgreater50 $\mu$s & \begin{tabular}[c]{@{}c@{}}Neglected for simulation simplicity,\\ $\ket{\downarrow}\rightarrow e^{i\delta t}\ket{\downarrow}$, $\ket{\uparrow}\rightarrow e^{-i\delta t}\ket{\uparrow}$,\\ where $\delta=2\pi\times 5$ kHz\end{tabular} \\ \hline
\multirow{4}{*}{\begin{tabular}[c]{@{}c@{}}Atomic thermal motion (80 $\mu$K), Ref.~\cite{Maller2015} \\$\delta\Omega < 0.01\Omega$, $\delta V\sim 0.5V$ \\Doppler shift\end{tabular}} &  & &  \\
 &  & \multicolumn{1}{c}{-} & Monte-Carlo method \\
 & \multirow{2}{*}{} & \multirow{2}{*}{  $\simeq$100 $\mu$s} & \multirow{2}{*}{\begin{tabular}[c]{@{}c@{}}Neglected for simulation simplicity\end{tabular}} \\
  \\ \hline
\end{tabular}
\end{table*}
\begin{table*}[tbh]
\centering
\caption{Global dephasing sources}
\label{T.collectivedephasing}
\begin{tabular}{cccc}
\hline
\multicolumn{2}{c}{Error sources} & Effective Rabi decay $\tau$ & Treatment \\ \hline
\multicolumn{2}{c}{\multirow{4}{*}{Rydbeg Lasers linewidth, $\leq$10 kHz}} & \multirow{4}{*}{$\geq$20 $\mu$s} & \multirow{4}{*}{\begin{tabular}[c]{@{}c@{}}$L_c=\sum_i \sqrt{\gamma_c/2}\sigma_z^i$,\\ where $\gamma\leq 2\pi\times 16$ kHz\end{tabular}} \\
\multicolumn{2}{c}{} &  &  \\
\multicolumn{2}{c}{} &  &  \\
\multicolumn{2}{c}{} &  &  \\ \hline
\begin{tabular}[c]{@{}c@{}}480 nm and dipole trap \\ beam pointing fluctuation\end{tabular} & \begin{tabular}[c]{@{}c@{}}$\delta\Omega\sim 0.05\Omega$,\\ $\delta\Delta\sim 2\pi\times 10$ kHz\end{tabular} & \multirow{4}{*}{$\simeq$2 $\mu$s} & \multirow{4}{*}{\begin{tabular}[c]{@{}c@{}}Monte-Carlo method  \end{tabular}} \\ \cline{1-2}
\begin{tabular}[c]{@{}c@{}}Intensity fluctuation,\\ 4\% (480), 2\% (780)\end{tabular} & \begin{tabular}[c]{@{}c@{}}$\delta\Omega\sim 0.03\Omega$,\\ $\delta\Delta\sim 2\pi\times 40$ kHz\end{tabular} &  &  \\ \cline{1-2}
\multirow{2}{*}{Static electric field fluctuation} & \multirow{2}{*}{$\delta\Delta\sim 2\pi\times 50$ kHz} &  &  \\
 &  &  &  \\ \hline
\end{tabular}
\end{table*}

\begin{figure}[tbh]
\includegraphics[width=0.46\textwidth]{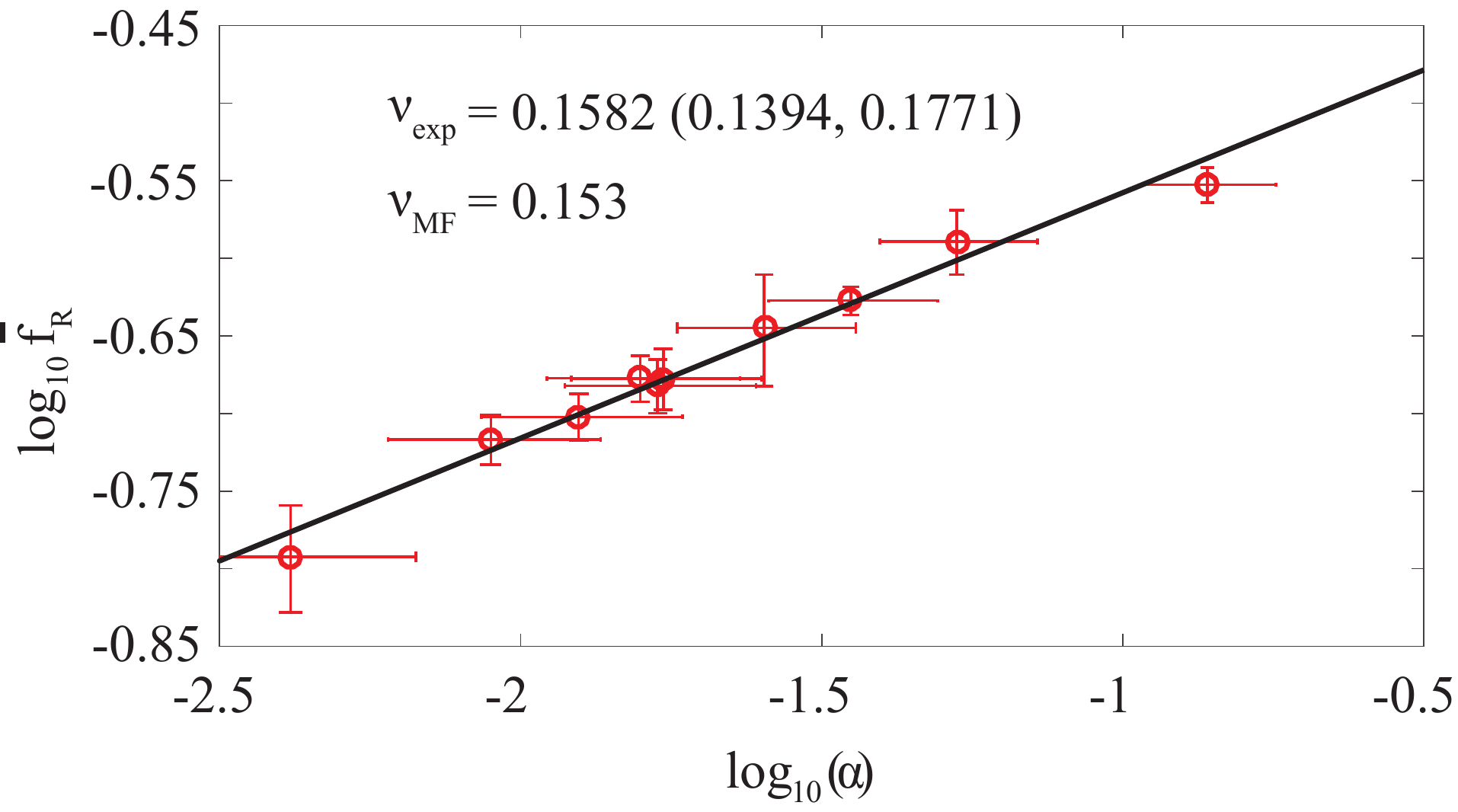} 
\caption{Scaling behavior of  $\bar{f}_R\sim\alpha^\nu$ as a function of $\alpha$ in the chains of $N=15$, where $\alpha$ is a function of $\theta$. 
The experimentally measured  exponent $\nu_{exp}$ is estimated with 95~\% confidence interval of least-square fit.
The y-axis (x-axis) errorbars indicate the standard deviation of $\bar{f}_R$ (the error by  $\sim 200$~nm fluctuation of $d$).}
\label{figsCriticalBehaviors}
\end{figure}
It was theoretically predicted~\cite{Low2012, Weimer2008} that in the atom chain described by the Hamiltonian $H$,
$\bar{f}_R$ follows the universal scaling behavior of $\bar{f}_R \propto \alpha^\nu$ with $\alpha = \hbar \Omega   / (|C_6| n_{eff}^6)$. $n_{eff}$ is the one-dimensional density of atoms.
For a zigzag chan, the density depends on the interatom distance $d$ and the bending angle $\theta$.
We define $n_{eff}$ as  $n_{eff}(\theta)\equiv n_{\parallel}/2+\min{(n_{\parallel}/2, n_{\perp})}$, where
 $n_{\parallel}\equiv(d\sin \frac{\theta}{2})^{-1}$ and $n_{\perp}\equiv(d\cos \frac{\theta}{2})^{-1}$;
 the definition connects the two limiting cases,
  $n_{eff}(\theta)=n_{\parallel}$ for $\theta\rightarrow180^\circ$ and $n_{eff}(\theta)=n_{\parallel}/2$ for $\theta\rightarrow 0^\circ$.
 One can choose other proper definitions of $n_{eff}$ for the zigzag chain. We emphasize that the scaling exponent $\nu$ obtained in our experiment is independent of the choice.
 

In Fig.~\ref{figsCriticalBehaviors}, we plot $\bar{f}_\textrm{R}$ as a function of $\alpha$. Here, $\alpha$ changes with $\theta$; in our experiment, $d = 4.0(2)$ $\mu$m is fixed, but $\theta$ is tuned.
The measured exponent $\nu_\textrm{exp} \simeq 0.158$ agrees with the prediction $\nu_\textrm{MF} = 0.153$~\cite{Low2012, Weimer2008}.
This supports that our system is well described by the Hamiltonian $H$.

\subsection{Error estimation and Lindblad equation}

\begin{figure*} [tbh]
\centering
\includegraphics[width=0.96\textwidth]{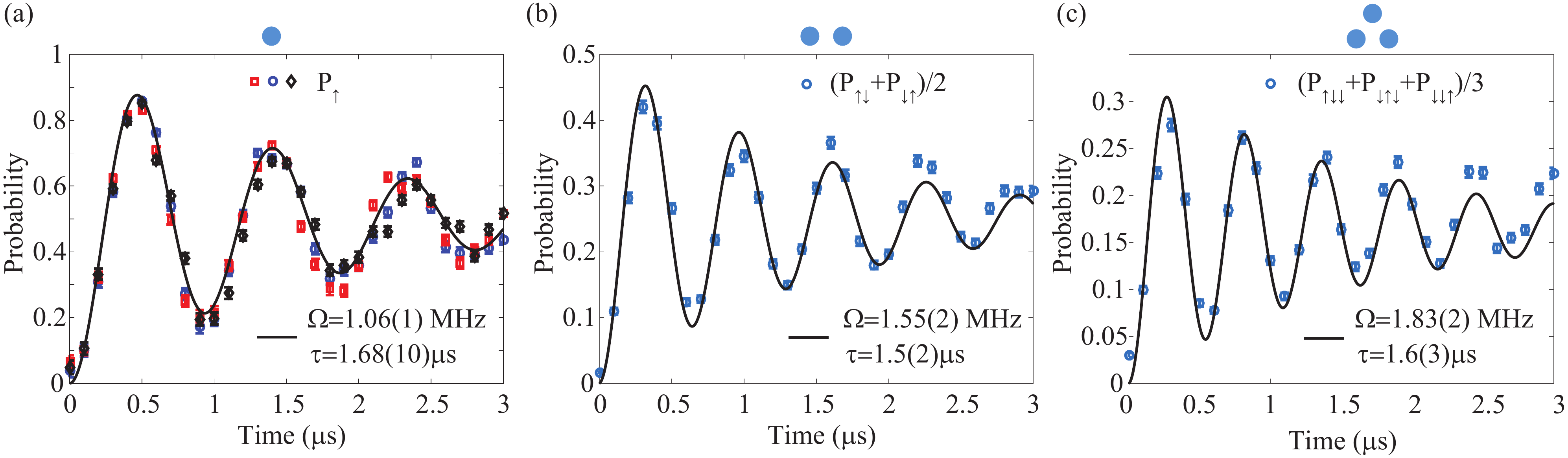} 
\caption{{\bf{Collective Rabi oscillations.}} Rabi oscillations of a short $\theta = 60^\circ$ zigzag chain.
The points represent the experimental data and the solid lines are the calculation results based on the Lindblad equation and the Monte-Carlo method.
(a) The Rabi oscillation of the probability $P_\uparrow$ of finding the atom in spin up in the case of $N=1$.
(b) The probability of finding one atom in spin up and the other atom in spin down in the case of $N=2$.
(c) The probability of finding one atom in spin up and the other two atoms in spin down in the case of $N=3$.
All the results are well fitted by $\{1-\cos(\Omega t)\exp(-t/\tau)\}/2N$ with the values of $\Omega$ and $\tau$ indicated in the figures.
In  (a), the color points are experimental data obtained from different positions (4.4~$\mu$m apart from each other) of the atom trap. }
\label{figs_RyfractionData}
\end{figure*}

The possible sources of errors in our quantum simulator are summarized in Tables~\ref{T.individualdephasing} and \ref{T.collectivedephasing}. To estimate the errors~\cite{Tomography1,Tomography2}, we combine a Lindblad equation and the Monte-Carlo method. The Lindblad equation was constructed~\cite{Lindblad1,Lindblad2} based on the information in Fig.~\ref{figs_levelscheme}(b),
\begin{equation}
\frac{d}{dt}\rho=-\frac{i}{\hbar}[H,\rho]+\mathcal{L}(\rho),
\label{eq.lindblad}
\end{equation}
where $\mathcal{L}(\rho)$ is Lindblad super-operator,
\begin{eqnarray}
\mathcal{L}(\rho)  & = & \sum_i L_i\rho L_i^\dagger-(L_i^\dagger L_i \rho+\rho L_i^\dagger L_i)/2 \nonumber \\
& + & L_c \rho L_c^\dagger -(L_c^\dagger L_c  \rho+\rho L_c^\dagger L_c)/2,
\label{eq.lindblad2}
\end{eqnarray}
$L_i = \sigma_z^i \sqrt{\gamma / 2}$ describes dephasing sources (such as spontaneous decay) on indivial atoms,
 $L_c = \sum_i \sigma_z^i \sqrt{\gamma_c / 2}$ describes golobal dephasing sources (such as laser linewidth), and $\gamma$ and $\gamma_c$ are parameters determined from the information in Fig.~\ref{figs_levelscheme}(b).
The shot-to-shot fluctuation of the parameters $\Omega$ and $\Delta$ of the Hamiltonian $H$ was taken into account by using the Monte-Carlo method in which the fluctuation is assumed to follow the Lorentizian distribution functions of 
$P(\Omega)\propto 1/(1+(\Omega-\Omega_0)^2/\delta\Omega^2)$ and $P(\Delta)\propto 1/(1+(\Delta-\Delta_0)^2/\delta\Delta^2)$. The parameters were determined by minimizing a maximum-likelihood estimation with the experimental data in Fig.~\ref{figs_RyfractionData}(a), where $\Omega_0=1.04$~MHz, $\delta\Omega=0.08$~MHz, $\Delta_0=0$, and $\delta\Delta=0.1$~MHz, respectively. Following the above steps, we obtain the solid line in Fig.~\ref{fig2}(a).

To analyze the errors, we first apply the above method to the short zigzag chains of $N=1,2,3$ and $\theta = 60^\circ$.
The calculation results based on the parameters determined above agree with the experimental data of Rabi oscillations in Fig.~\ref{figs_RyfractionData}(b,c). 
The Rabi oscillations are well fitted by the exponential decay of  $\{1-\cos(\Omega t)\exp(-t/\tau)\}/2N$.
The decay time $\tau$ is almost the same for $N=1,2,3$.
Our analysis indicates that  more than 80~\% of the Rabi oscillation decay in our experiment was due to inhomogeneous dephasing from slow environmental fluctuations such as beam power, stray electric field, and beam pointing fluctuation. See Tables~\ref{T.individualdephasing} and \ref{T.collectivedephasing}.

\subsection{Implementation details of Matrix Product States}


We used the matrix product states (MPS)~\cite{Schollwock2011} of  
\begin{align}
|\psi\rangle
= \sum_{\sigma_z^{(1)},\cdots,\sigma_z^{(L)}} M^{\sigma_z^{(1)}} M^{\sigma_z^{(2)}}\cdots M^{\sigma_z^{(L)}}|\sigma_z^{(1)}, \sigma_z^{(2)}, \cdots, \sigma_z^{(L)}\rangle.
\end{align}
Here, $\sigma_z^{(i)} = 1$ ($\sigma_z^{(i)} = -1$) represents pseudospin $\uparrow_i$ ($\downarrow_i$).
$\sigma_z^{(i)} = \pm 1$ is related to the occupation number $n_i$ of the Rydberg state of atom $i$ through the relation $\hat{n}_i = (1+\hat{\sigma}_i^z)/2$. $M^{\sigma_z^{(i)}}$ is a $(\chi_i \times \chi_{i+1})$ matrix for each quantum number $\sigma_z^{(i)}$. The integer $\chi_i$ is called the bond dimension.  

The initial state at $t=0$ is $|\psi_0\rangle=|-1,-1,\dots, -1\rangle$. It is a MPS with bond dimension $\chi_j = 1$ for all $j$. We evolve the initial state using the Hamiltonian 
\begin{align*}
H = V_{12} \sum_{i=1}^{N-1} \hat{n}_i \hat{n}_{i+1} + V_{13} \sum_{i=1}^{N-2} \hat{n}_i \hat{n}_{i+2} +\frac{\Omega}{2} \sum_{i=1}^N  \hat{\sigma}_x^{(i)},
\end{align*} 
where we choose $V_{12}/2\pi \hbar=20$~MHz, $V_{13} = V_{12} / 64$ for $\theta = 180^\circ$,
$V_{13} = V_{12}$ for $\theta = 60^\circ$, and $\Omega/2\pi=1.0(1)$~MHz. 
We use the second order Suzuki-Trotter approximation,
\begin{align}
e^{-i H dt} \approx e^{-ih_x dt/2} e^{-ih_z dt} e^{-i h_x dt/2} ,
\end{align}
where $h_x$ ($h_z$) is the part of $H$ containing $\sigma_x^{(i)}$'s ($\sigma_z^{(i)}$'s).
 The operator $h_z$ contains $\sum_{i=1}^{L-1}\sigma_z^{(i)} \sigma_z^{(i+1)}$ and $\sum_{i=1}^{L-2}\sigma_z^{(i)} \sigma_z^{(i+2)}$, and their exponentials can be written as  matrix product operators (MPO) using the method described in Ref.~\cite{Pirvu2010}.
We apply the MPO of $e^{-iHdt}$ on $|\psi_0\rangle$. Because this operation increases bond dimensions, we variationally compress~\cite{Verstraete2004,Verstraete2008} the resulting state. In other words, we approximate the resulting state with another MPS with bond dimensions smaller than some fixed maximum bond dimension $\chi_\text{max}$. We repeat this process  to evolve $|\psi_0\rangle$ over finite time $t$.

There are two sources of error: (1) the time step in Suzuki-Trotter approximation is not infinitesimal and (2) the maximum bond dimension is finite. We verified that the time step smaller than $\Omega dt = 0.013$ does not change our data much. We increased $\chi_\text{max}$ until our data shows no appreciable dependence on $\chi_\text{max}$. For example, when $ V_{13}=0$ and $N = 23$, we had $\chi_\text{max} = 240$. When  $ V_{13}=V_{12}$ and $N=26$, we had $\chi_\text{max} = 200$.


\subsection{Invalidity of ETH in our cases} 

If the eigenstate thermalization hypothesis (ETH) holds for our system, the matrix elements of $\hat n_i$ in the basis of the eigenstates $|\alpha\rangle$ of $ H$ can be written as~\cite{Srednicki1999, D'Alessio2016} 
\begin{align*}
n_{\alpha\beta}(E,\omega) \equiv \langle \alpha| \hat n_i |\beta \rangle  
= n(E)\, \delta_{\alpha\beta} + e^{-S(E)/2} f(E, \omega) R_{\alpha\beta},
\end{align*}
where $ H |\alpha \rangle = E_\alpha | \alpha \rangle$, $E = (E_\alpha + E_\beta)/2$, and $\omega = E_\alpha- E_\beta$.  $n(E)$ and $f(E,\omega)$ are smooth functions of $E$ and $\omega$. As our system is time-reversal-symmetric, $R_{\alpha\beta} = R_{\beta \alpha}$ is real, and $R_{\alpha\beta}$ is a random variable with zero mean and unit variance. $S(E)$ is the thermodynamic entropy at $E$.

\begin{figure}[tbh]
\includegraphics[width=0.92\columnwidth]{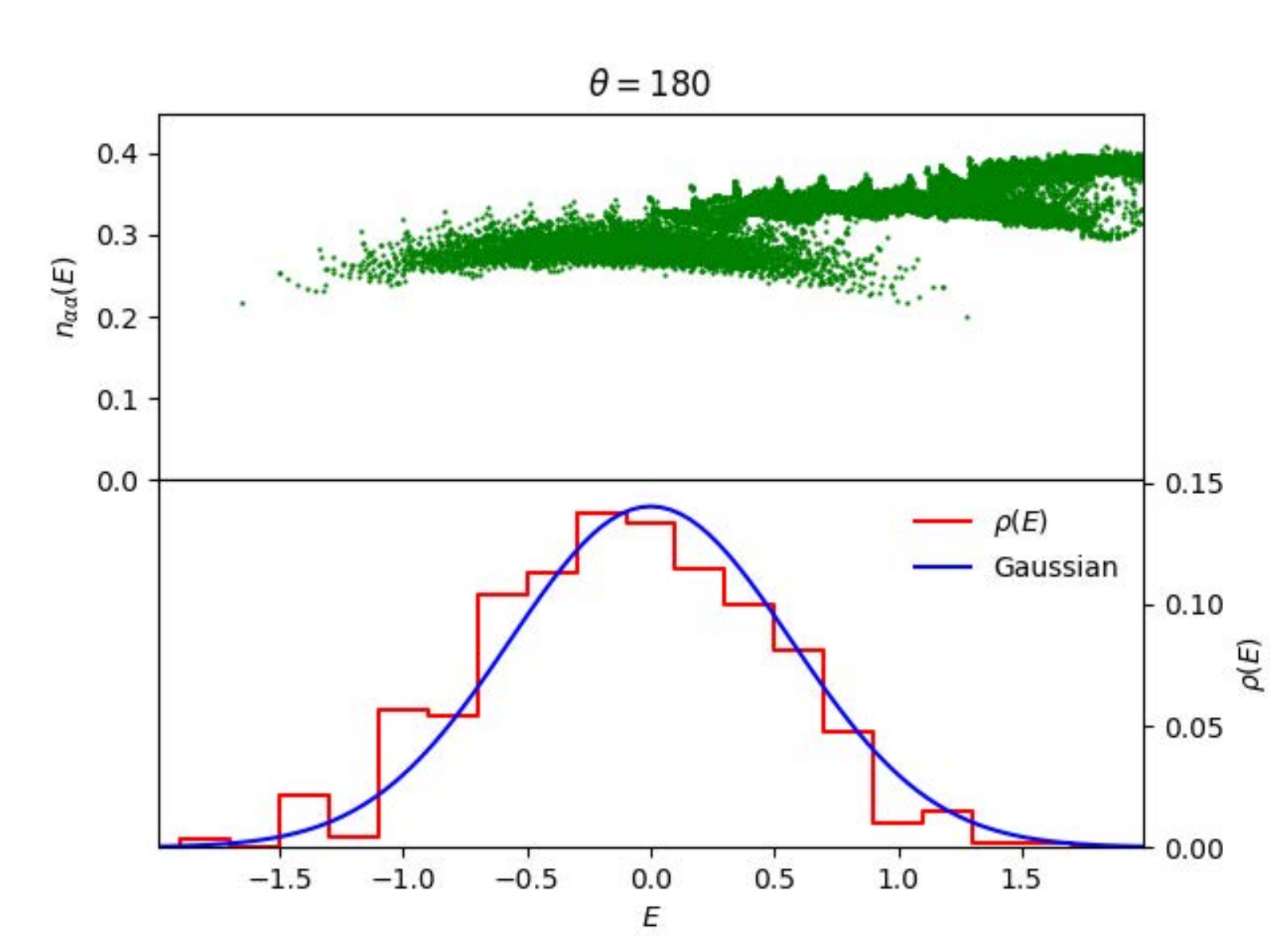}
\caption{$n_{\alpha \alpha} (E)$ and $\rho(E)$ for the linear chain with $N=19$.  
Upper panel: The diagonal element $n(E)$ is supposed to be a smooth function of $E$ and constant over energy windows with width $\Delta$, according to ETH. However, we see that $n(E)$ is not smooth. 
$\rho(E)$ is plotted after summing delta functions over small energy windows. 
If the system is quantum chaotic, $\rho(E)$ is expected to have a Gaussian centered at 0 with standard deviation $\Delta = 0.57$.
} \label{theta180}
 
\includegraphics[width=0.92\columnwidth]{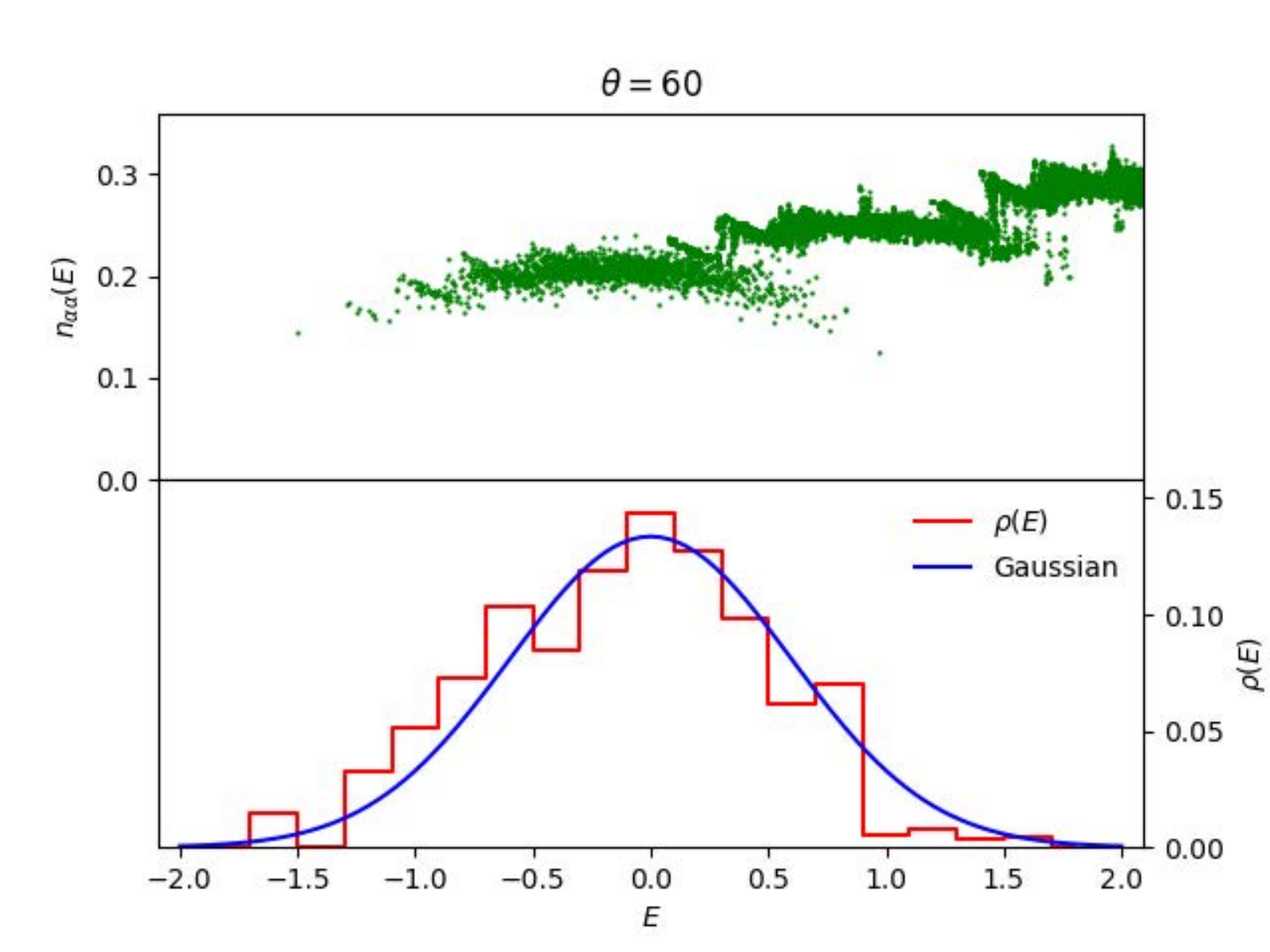}
\caption{$n_{\alpha \alpha} (E)$ and $\rho(E)$ for the zigzag chain with $N=21$ and $\theta = 60^\circ$.  
The diagonal element $n(E)$ is supposed to be a smooth function of $E$ and constant over energy windows with width $\Delta$, according to ETH. However, we see that $n(E)$ is not smooth. 
$\rho(E)$ is plotted after summing delta functions over small energy windows. 
If the system is quantum chaotic, $\rho(E)$ is expected to have a Gaussian centered at 0 with standard deviation $\Delta = 0.6$.
} \label{theta60}
\end{figure}

To see whether the above feature is satisfied in our case, we plot the diagonal elements $n_{\alpha\alpha}(E) = \langle \alpha | \hat n_i |\alpha \rangle$ versus $E = E_\alpha$ for a linear chain in Fig.~\ref{theta180} and for a zigzag chain in Fig.~\ref{theta60}; similar figures have been studied in Ref.~\cite{Rigol2008}. If the ETH is correct, $n_{\alpha\alpha}(E)$ should be smooth and nearly constant within the energy window $[-\Delta,\Delta]$ as the initial state $|\downarrow_1 \downarrow_2 \cdots \downarrow_N \rangle$ has the energy expectation value $\langle  H \rangle = 0$ and the energy fluctuation $\Delta =  \sqrt{\langle  H^2 \rangle - \langle  H \rangle^2} = \Omega \sqrt N$.
However, $n_{\alpha\alpha}(E)$ is not smooth over $E_\alpha\in [-\Delta,\Delta]$ in our cases shown in Figs.~\ref{theta180} and \ref{theta60}, demonstrating the violation of the ETH.
  
Note that we also plot the normalized energy distribution $\rho(E)$ 
\begin{align*}
\rho(E) = \sum_\alpha |C_\alpha|^2 \delta(E- E_\alpha)
\end{align*}
in Figs.~\ref{theta180} and \ref{theta60}, where $C_\alpha = \langle \alpha |\downarrow_1 \downarrow_2 \cdots \downarrow_N \rangle$.
The distribution tells us which energy eigenstates give the most dominant contributions to the eigenstate expansion of $ |\downarrow_1 \downarrow_2 \cdots \downarrow_N \rangle = \sum_\alpha C_\alpha |\alpha \rangle$. 
$\rho(E)$ is almost a Gaussian with mean 0 and standard deviation $\Delta$. 

\noindent \textbf{Ackowledgements}

This research was supported by the Samsung Science and Technology Foundation [SSTF-BA1301-12] and Korea National Research Foundation  [2016R1A5A1008184].
 

\begin{thebibliography}{11}
\bibitem{Pathria} Pathria, R. K. \& Beale, P. D. {\textit{Statistical Mechanics.}}  (Elsevier Science, 1996).

\bibitem{Polkovnikov2011} Polkovnikov, A., Sengupta, K., Silva, A. \& Vengalattore, M. Colloquium: Nonequilibrium dynamics of closed interacting quantum systems. \textit{Rev. Mod. Phys.} \textbf{83}, 863-883 (2011).

\bibitem{Eisert2015} Eisert, J., Friesdorf, M. \& Gogolin, C. Quantum many-body systems out of equilibrium. \textit{Nat Phys} \textbf{11}, 124-130, doi:10.1038/nphys3215 (2015).

\bibitem{Christian2016} Christian, G. \& Jens, E. Equilibration, thermalisation, and the emergence of statistical mechanics in closed quantum systems. \textit{Rep. Prog. Phys.} \textbf{79}, 056001 (2016).


\bibitem{Langen2013} Langen, T., Geiger, R., Kuhnert, M., Rauer, B. \& Schmiedmayer, J. Local emergence of thermal correlations in an isolated quantum many-body system. \textit{Nat Phys} \textbf{9}, 640-643 (2013).

\bibitem{Clos2016} Clos, G., Porras, D., Warring, U. \& Schaetz, T. Time-Resolved Observation of Thermalization in an Isolated Quantum System. \textit{Phys. Rev. Lett.} \textbf{117}, 170401 (2016).

\bibitem{Georg2016} Georg, K. \textit{et al}. Critical thermalization of a disordered dipolar spin system in diamond. \textit{arXiv}: 1609.08216 (2016).

\bibitem{Kaufman2016} Kaufman, A. M. \textit{et al}. Quantum thermalization through entanglement in an isolated many-body system. \textit{Science} \textbf{353}, 794-800 (2016).

\bibitem{Neill2016} Neill, C.\textit{ et al}. Ergodic dynamics and thermalization in an isolated quantum system. \textit{Nat Phys} \textbf{12}, 1037-1041, (2016).

\bibitem{Deutsch1991} Deutsch, J. M. Quantum statistical mechanics in a closed system. \textit{Physical Review A} \textbf{43}, 2046-2049 (1991).

\bibitem{Srednicki1994} Srednicki, M. Chaos and quantum thermalization. \textit{Phys Rev E} \textbf{50}, 888-901 (1994).

[srednicki1999]
\bibitem{Srednicki1999} Mark, S. The approach to thermal equilibrium in quantized chaotic systems. \textit{J. Phys. A: Math. Gen.} \textbf{32}, 1163 (1999).

\bibitem{Rigol2008} Rigol, M., Dunjko, V. \& Olshanii, M. Thermalization and its mechanism for generic isolated quantum systems. \textit{Nature} \textbf{452}, 854-858 (2008).

\bibitem{Rigol2012} Rigol, M. \& Srednicki, M. Alternatives to Eigenstate Thermalization. \textit{Phys. Rev. Lett.} \textbf{108}, 110601 (2012).

\bibitem{D'Alessio2016} D'Alessio, L., Kafri, Y., Polkovnikov, A. \& Rigol, M. From quantum chaos and eigenstate thermalization to statistical mechanics and thermodynamics. \textit{Adv. Phys.} \textbf{65}, 239-362 (2016).

\bibitem{Ates2012} Ates, C., Garrahan, J. P. \& Lesanovsky, I. Thermalization of a Strongly Interacting Closed Spin System: From Coherent Many-Body Dynamics to a Fokker-Planck Equation. \textit{Phys. Rev. Lett.} \textbf{108}, 110603 (2012).


\bibitem{Kim2016} Kim, H.\textit{ et al}. In situ single-atom array synthesis using dynamic holographic optical tweezers. \textit{Nature Communications} \textbf{7}, 13317 (2016).

\bibitem{Lee2017} Lee, W., Kim, H. \& Ahn, J. Defect-free atomic array formation using the Hungarian matching algorithm. \textit{Physical Review A} \textbf{95}, 053424 (2017).

\bibitem{Barredo2016} Barredo, D., de L\'eés\'eéleuc, S., Lienhard, V., Lahaye, T. \& Browaeys, A. An atom-by-atom assembler of defect-free arbitrary two-dimensional atomic arrays. \textit{Science} \textbf{354}, 1021-1023 (2016).

\bibitem{Endres2016} Endres, M.\textit{ et al}. Atom-by-atom assembly of defect-free one-dimensional cold atom arrays. \textit{Science} \textbf{354}, 1024-1027 (2016).


\bibitem{Low2012} Low, R.\textit{ et al}. An experimental and theoretical guide to strongly interacting Rydberg gases. \textit{J Phys B-at Mol Opt} \textbf{45}, 113001 (2012).

\bibitem{Schaub2015} Schau\ss{}, P.\textit{ et al}. Crystallization in Ising quantum magnets. \textit{Science} \textbf{347}, 1455-1458 (2015).

\bibitem{Labuhn2016} Labuhn, H.\textit{ et al}. Tunable two-dimensional arrays of single Rydberg atoms for realizing quantum Ising models. \textit{Nature} \textbf{534}, 667-670 (2016).

\bibitem{Bernien2017} Bernien, H.\textit{ et al}. Probing many-body dynamics on a 51-atom quantum simulator. \textit{arXiv}:1707.04344v1 (2017).

\bibitem{Lienhard2017} Lienhard, V.\textit{ et al}. Observing the space- and time-dependent growth of correlations in dynamically tuned synthetic Ising antiferromagnets. \textit{arXiv}:1711.01185 (2017).

\bibitem{Sebastian2017} Sebastian, W.\textit{ et al}. Calculation of Rydberg interaction potentials. \textit{J. Phys. B: At. Mol. Phys.} \textbf{50}, 133001 (2017).


\bibitem{Lesanovsky2010} Lesanovsky, I., Olmos, B. \& Garrahan, J. P. Thermalization in a Coherently Driven Ensemble of Two-Level Systems. \textit{Phys. Rev. Lett.} \textbf{105}, 100603 (2010).

\bibitem{Kiendl2017} Kiendl, T. \& Marquardt, F. Many-Particle Dephasing after a Quench. \textit{Phys. Rev. Lett.} \textbf{118}, 130601 (2017).

\bibitem{Weimer2008} Weimer, H., L\"oöw, R., Pfau, T. \& B\"uüchler, H. P. Quantum Critical Behavior in Strongly Interacting Rydberg Gases. \textit{Phys. Rev. Lett.} \textbf{101}, 250601 (2008).


\bibitem{Marcuzzi2017} Marcuzzi, M.\textit{ et al}. Facilitation Dynamics and Localization Phenomena in Rydberg Lattice Gases with Position Disorder. \textit{Phys. Rev. Lett.} \textbf{118}, 063606 (2017).

\bibitem{Guardado2017} Guardado-Sanchez, E.\textit{ et al}. Probing quench dynamics across a quantum phase transition into a 2D Ising antiferromagnet. \textit{arXiv}:1711.00887 (2017).


\bibitem{Olmos2010} Olmos, B., M\"uüller, M. \& Lesanovsky, I. Thermalization of a strongly interacting 1D Rydberg lattice gas. \textit{New J. Phys.} \textbf{12}, 013024 (2010).

\bibitem{Maller2015} Maller, K. M.\textit{ et al}. Rydberg-blockade controlled-not gate and entanglement in a two-dimensional array of neutral-atom qubits. \textit{Physical Review A} \textbf{92}, 022336 (2015).

\bibitem{Tomography1} Zhang, X. L., Gill, A. T., Isenhower, L., Walker, T. G. \& Saffman, M. Fidelity of a Rydberg-blockade quantum gate from simulated quantum process tomography. \textit{Physical Review A} \textbf{85}, 042310 (2012).

\bibitem{Tomography2} Miroshnychenko, Y.\textit{ et al}. Coherent excitation of a single atom to a Rydberg state. \textit{Physical Review A} \textbf{82}, 013405 (2010).

\bibitem{Lindblad1} Lindblad, G. On the generators of quantum dynamical semigroups. \textit{Comm. Math. Phys.} \textbf{48}, 119-130 (1976).

\bibitem{Lindblad2} Gorini, V., Kossakowski, A. \& Sudarshan, E. C. G. Completely positive dynamical semigroups of N‐level systems. \textit{J Math Phys} \textbf{17}, 821-825 (1976).

\bibitem{MethodsBeterov} Beterov, I. I., Ryabtsev, I. I., Tretyakov, D. B. \& Entin, V. M. Quasiclassical calculations of blackbody-radiation-induced depopulation rates and effective lifetimes of Rydberg nS, nP, and nD alkali-metal atoms with n $\leq$ 80. \textit{Physical Review A} \textbf{79}, 052504 (2009).
 
\noindent
\bibitem{Schollwock2011} Schollw\"oöck, U. The density-matrix renormalization group. \textit{Rev. Mod. Phys.} \textbf{77}, 259-315 (2005).

\noindent
\bibitem{Pirvu2010} Pirvu, B., Murg, V., Cirac, J. I. \& Verstraete, F. Matrix product operator representations. \textit{New J. Phys.} \textbf{12}, 025012 (2010).

\noindent
\bibitem{Verstraete2004} Verstraete, F., Garc\'iía-Ripoll, J. J. \& Cirac, J. I. Matrix Product Density Operators: Simulation of Finite-Temperature and Dissipative Systems. \textit{Phys. Rev. Lett.} \textbf{93}, 207204 (2004).

\noindent
\bibitem{Verstraete2008} Verstraete, F., Murg, V. \& Cirac, J. I. Matrix product states, projected entangled pair states, and variational renormalization group methods for quantum spin systems. \textit{Adv. Phys.} \textbf{57}, 143-224 (2008).

\end{thebibliography}
\end{document}